\documentclass[final]{siamltex1213}

\usepackage{graphicx,amsmath,color,listings}

\usepackage[symbol*]{footmisc}

\title{MADNESS: A Multiresolution, Adaptive Numerical Environment for Scientific Simulation \thanks{Corresponding author: Robert J.\ Harrison (\texttt{robert.harrison@stonybrook.edu}). This research was sponsored in part by the Office of Advanced Scientific Computing Research of the U.S.\ Department of Energy, and includes work performed at the Oak Ridge National Laboratory, which is managed by UT-Battelle, LLC under Contract No.\ DE-AC05-00OR22725. This work employed resources of the National Center for Computational Sciences at Oak Ridge National Laboratory and of the Argonne Leadership Computing Facility, which is a DOE Office of Science User Facility supported under Contract DE-AC02-06CH11357. R.J.H.\ was supported in part by the National Science Foundation under grant ACI-1450344.}}

\renewcommand{\thefootnote}{\arabic{footnote}}

\newcommand*{\sbu}{2}
\newcommand*{\ucb}{3}
\newcommand*{\humboldt}{4}
\newcommand*{\vtu}{5}
\newcommand*{\ornl}{6}
\newcommand*{\fsu}{7}
\newcommand*{\intel}{8}
\newcommand*{\lbnl}{9}
\newcommand*{\linkedin}{10}
\newcommand*{\ud}{11}
\newcommand*{\anl}{12}
\newcommand*{\uw}{13}
\newcommand*{\tut}{14}
\newcommand*{\lsu}{15}
\newcommand*{\lasierra}{16}

\author{Robert~J. Harrison\footnotemark[\sbu]
\and Gregory Beylkin\footnotemark[\ucb]
\and Florian~A. Bischoff\footnotemark[\humboldt]
\and Justus~A. Calvin\footnotemark[\vtu]
\and George~I. Fann\footnotemark[\ornl]
\and Jacob Fosso-Tande\footnotemark[\fsu]
\and Diego Galindo\footnotemark[\ornl]
\and Jeff~R. Hammond\footnotemark[\intel]
\and Rebecca Hartman-Baker\footnotemark[\lbnl]
\and Judith~C. Hill\footnotemark[\ornl]
\and Jun Jia\footnotemark[\linkedin]
\and Jakob~S. Kottmann\footnotemark[\humboldt]
\and M-J.~Yvonne Ou\footnotemark[\ud]
\and Laura~E. Ratcliff\footnotemark[\anl]
\and Matthew~G. Reuter\footnotemark[\sbu]
\and Adam~C. Richie-Halford\footnotemark[\uw]
\and Nichols~A. Romero\footnotemark[\anl]
\and Hideo Sekino\footnotemark[\tut]
\and William~A. Shelton\footnotemark[\lsu]
\and Bryan~E. Sundahl\footnotemark[\sbu]
\and W.~Scott Thornton\footnotemark[\sbu]
\and Edward~F. Valeev\footnotemark[\vtu]
\and \'Alvaro V\'azquez-Mayagoitia\footnotemark[\anl]
\and Nicholas Vence\footnotemark[\lasierra]
\and Yukina Yokoi\footnotemark[\tut]
}

\begin{document}
\maketitle

\renewcommand{\thefootnote}{\fnsymbol{footnote}}

\footnotetext[\sbu]{Stony Brook University, Stony Brook, New York 11794, USA}
\footnotetext[\ucb]{University of Colorado at Boulder, Boulder, Colorado 80309, USA}
\footnotetext[\humboldt]{Institut f{\"u}r Chemie, Humboldt-Universit{\"a}t zu Berlin, Unter den Linden 6, 10099 Berlin, Germany}
\footnotetext[\vtu]{Virginia Tech, Blacksburg, Virginia 24061, USA}
\footnotetext[\ornl]{Oak Ridge National Laboratory, Oak Ridge, Tennessee 37831, USA}
\footnotetext[\fsu]{Department of Chemistry and Biochemistry, Florida State University, Tallahassee, Florida 32306, USA}
\footnotetext[\intel]{Parallel Computing Lab, Intel Corporation, Portland, Oregon 97219, USA}
\footnotetext[\lbnl]{National Energy Research Scientific Computing Center, Lawrence Berkeley National Laboratory, Berkeley, California 94720, USA}
\footnotetext[\linkedin]{LinkedIn Corporation, Mountain View, California 94043, USA}
\footnotetext[\ud]{Department of Mathematical Sciences, University of Delaware, Newark, Delaware 19716, USA}
\footnotetext[\anl]{Argonne Leadership Computing Facility, Argonne National Laboratory, Argonne, Illinois 60439, USA}
\footnotetext[\uw]{Department of Physics, University of Washington, Seattle, Washington 98195, USA}
\footnotetext[\tut]{Computer Science and Engineering, Toyohashi University of Technology, Toyohashi, AICHI 441-8580, Japan}
\footnotetext[\lsu]{Louisiana State University, Baton Rouge, Louisiana 70803, USA}
\footnotetext[\lasierra]{Department of Physics, LaSierra University, Riverside, California 92505, USA}

\renewcommand{\thefootnote}{\arabic{footnote}}


\pagestyle{myheadings}
\thispagestyle{plain}
\markboth{R.~J. HARRISON ET AL.}{AN OVERVIEW OF MADNESS} 

\begin{abstract}
MADNESS
(\underline{m}ultiresolution \underline{ad}aptive \underline{n}umerical \underline{e}nvironment
for \underline{s}cientific \underline{s}imulation) is a high-level
software environment for solving integral and differential
equations in many dimensions that uses adaptive and fast harmonic analysis
methods with guaranteed precision based on multiresolution analysis
and separated representations.  Underpinning the numerical
capabilities is a powerful petascale parallel programming environment
that aims to increase both programmer productivity and code
scalability. This paper describes the features and capabilities of MADNESS
and briefly discusses some current applications 
in chemistry and several areas of physics.

\end{abstract}

\section{Introduction}
\label{sec:intro}
Numerical tools are ubiquitous in modern science since the relevant
models/equations do not generally have analytical solutions. Although
advances in scientific computing over the last thirty years have
enabled more sophisticated models and the quantitative simulation of
large-scale problems, these numerical methods introduce new,
unphysical considerations that must also be taken into account. For
instance, function representation, that is, the choice of basis set,
is critically important. If a basis set does not adequately resolve
the finest details of the system, the calculated solutions will likely
be inaccurate and have questionable physical interpretation.  Other
numerical factors include computational efficiency, scalability,
distributed memory management,
differentiation schemes, and quadrature rules.  As such, most
computational approaches force the scientist or engineer to compose
software in terms of details inherent to the representation (e.g.,
the coefficients of basis functions or integrals within the basis set),
rather than at the level of mathematics or physics that uses the natural
functions or operators of the problem domain. In the end, all of these
considerations obscure the desired science and force scientists to
instead focus on computational mundanities.

Several software packages have been developed to help insulate
scientists from these issues; two notable examples are
Trilinos \cite{trilinos-article} and PETSc \cite{petsc-user-ref}. In
essence, these and similar frameworks support common scientific computing
operations, \textit{e.g.}, linear algebra algorithms, mesh generation
(function representation), nonlinear solvers, and ordinary
differential equation integrators. These packages also typically scale to parallel or
massively-parallel machines, facilitating the solution of large-scale
problems.  Unfortunately, to our knowledge, all of these packages
require the scientist to ``think'' in vectors and matrices, instead
of functions and operators. Moreover, with an emphasis on engineering
applications, computation is typically restricted to three or fewer dimensions.

Our goal in this communication is to motivate and overview the
MADNESS
(\underline{m}ulti\-resolution \underline{ad}aptive \underline{n}umerical \underline{e}nvironment
for \underline{s}cientific \underline{s}imulation) package, which
offers the following capabilities to users:
\begin{remunerate}

\item 
The illusion of basis set-free computation. Under this illusion,
the user computes with functions and operators instead of vectors and
matrices. Behind the scenes, functions and operators are expanded in
a multi-wavelet basis, where the number of basis functions is adapted
to guarantee the precision of each operation. As necessary, the
``mesh'', which is the support of the basis functions, is refined in
computationally-troublesome regions (perhaps those with fine details)
and coarsened in others, enabling the simulation of multi-scale
problems.

\item 
Fast and arbitrarily-accurate methods for solving differential and
integral equations in one to six dimensions (perhaps more), with specified
boundary conditions. These operations include, for example, linear
algebra, numerical differentiation/integration, and, perhaps most important,
integral convolution kernels, such as the Poisson equation Green's function.

\item
A parallel runtime that allows the user to compose algorithms as a
dynamically scheduled set of tasks that operate on objects in global
namespaces.

\item 
Algorithms and implementations that scale to massively-parallel computational resources,
thereby enabling the solution of large problems.
\end{remunerate}

We aim, by providing these tools, to raise the level of composition of
scientific applications, making it faster and easier to both (i)
construct robust and correct algorithms and (ii) calculate solutions
to new and existing problems. Although MADNESS's breakthrough initial
application was in computational chemistry \cite{harrison-11587, yanai-2866,
yanai-6680}, its framework more generally represents
functions, operators and boundary conditions. Thus, it is readily applicable
to describe and solve well-defined equations and boundary conditions.
Extensive applications include, to date, atomic and molecular
physics, electrostatics, materials science, nanoscience, nuclear physics,
solvation models, and graph theory \cite{fann-012080, fosso-tande-179,
jia-13, reuter-1, reuter-013426, sullivan-1838, vence-033403}.
We hope that this introduction to MADNESS conveys
its generality and encourages its application to new domains.

The layout of this communication is as follows.
Section \ref{sec:numerics} introduces the basics of
multiresolution analysis (MRA), which is the mathematical underpinning
of MADNESS, and how MRA facilitates scientific computing.  We then
provide and discuss, in \S\ref{sec:examples}, some simple examples of
programming in MADNESS that illustrate the absence of basis set
considerations and the ability to write code in terms of functions and
operators. Then, \S\ref{sec:computation} summarizes the computational
details of MADNESS, including the runtime
environment/framework and the methods behind
MADNESS's utilization of (massively-)parallel
architectures. Finally, \S\ref{sec:summary} summarizes this report and
describes future directions for MADNESS.

\section{Numerics: Mathematical Overview \& Capabilities}
\label{sec:numerics}

With the illusion of basis free computation, a MADNESS user
simply needs to specify the desired accuracy; from this, the numerical framework
provides fast computation with guaranteed accuracy (in up to 6
dimensions). This is accomplished through several key mechanisms:
\begin{remunerate}
\item The underlying representation is an orthonormal basis \cite{alpert-149};
functions are represented on an adaptive mesh with a discontinuous
spectral-element basis set. Within each adaptively-refined element the
basis functions are tensor products of Legendre polynomials.
Crucially, the application programmer only uses the function as a
single entity and never sees or manipulates the basis or mesh.
\item Guaranteed accuracy for each operation (\textit{e.g.},\
multiplication or addition of functions, application of an operator)
is accomplished through automatic refinement of the mesh using
criteria derived from MRA \cite{alpert-149}.
\item Fast and accurate computation is enabled through MRA that separates
behavior between length scales. MRA provides a sparse hierarchical
(tree) decomposition of functions and operators that enables automatic
refinement or coarsening of the mesh to meet the accuracy goal. MRA
also yields fast algorithms for physically important operators
\cite{alpert-149,beylkin-235,fann-161}.
\item Fast and accurate computation beyond one dimension is made feasible
through accurate and compact representations of operators as 
sums of separable kernels (usually Gaussians), which are created 
through quadrature or more advanced numerical techniques
\cite{beylkin-235,beylkin-131,harrison-103}.
\end{remunerate}
In the remainder of this section, we elaborate on these points.

\subsection{Underlying basis and adaptive mesh}

Inside MADNESS (using a 1D example here, for simplicity), the user's computational domain 
is mapped to $[0,1]$. This domain is repeatedly subdivided by factors of
two such that, at level $n$, there are $2^n$ boxes, each of size
$2^{-n}$. Within each box, we use the first $k$ Legendre polynomials
as the basis (or scaling functions in the language of
MRA \cite{alpert-246,alpert-149}). The polynomials are scaled and shifted to form
an orthonormal basis.

Specifically, the $i^\mathrm{th}$ scaling
function ($i=0,\ldots,k-1$) in the $l^\mathrm{th}$ box
($l=0,\ldots,2^{n}-1$) at level $n$ ($n=0,1,\ldots$) is
\[
\phi^n_{li}(x) = 2^{n/2} \phi_i(2^n x - l),
\]
where
\begin{equation}
\phi_i(x) = \left\{ \begin{array}{rl}
                        \sqrt{2i+1}\, P_i(2x-1), & x \in (0,1) \\
                         0, & x \not\in (0,1) 
                    \end{array} \right.
\label{eq:scalingfunc}
\end{equation}
is the $i^\mathrm{th}$ mother scaling function. Projection of a function $f(x)$
into the basis at level $n$ is accomplished by orthogonal projection,
\begin{subequations}
\begin{align}
   f^{n}(x) & = \sum_{li} s^n_{li} \phi^n_{li}(x) \\
    s^n_{li} & = \int \mathrm{d}x f(x) \phi^n_{li}(x)
\end{align}
\label{eq:projection}
\end{subequations}
Projection is not performed globally on a uniform fine grid; instead,
projection starts on a relatively coarse level ($n$). Within each box on
level $n$ we compute the projection at level $n$ and at the coarser level
$n-1$. If the difference between these projections is not satisfactory, the
projection is locally repeated at a finer level ($n+1$), and so forth until the
projection is sufficiently accurate. The differences in representations between
successive levels are efficiently and accurately computed in the
wavelet basis, which we now describe.

\subsection{Multiresolution analysis}

Rather than thinking about the approximation of a function at each
length scale, MRA focuses upon the difference between length scales
and, by doing so, exploits both smoothness and sparsity for fast
computation. The telescoping series clearly
illustrates this. Let $V^n$ be the space spanned by the polynomials
at level $n$. Then, we
can write
\begin{equation}
  V^n = V^0 \oplus \left( V^1  \ominus V^0 \right) \oplus \left( V^2  \ominus V^1 \right) \oplus \cdots \oplus \left( V^{n}  \ominus V^{n-1} \right),
\end{equation}
which decomposes the approximation space at level $n$ in terms of
the coarsest approximation ($V^0$) plus corrections at successive levels of
refinement. If the function of interest is sufficiently smooth in some volume, then
differences at finer scales will be negligible in that volume. Thus, smoothness is
translated into sparsity and hence fast computation; a precise
definition of ``negligible'' enables adaptive refinement to guarantee
accuracy. The multi-wavelets at level $n$ are defined as an
orthonormal, disjoint basis that spans the difference space $W^n =
V^{n+1} \ominus V^{n}$ and are constructed by translation and dilation
of the mother wavelets \cite{alpert-246,alpert-149}.

We thus have two representations of our function: (i) its projection
onto the finest level of the adaptively refined grid (\textit{i.e.},
in real space or in the scaling function basis) and (ii) its
representation at the coarsest level plus corrections at successive
levels of refinement (\textit{i.e.}, in the multi-wavelet basis). The
fast wavelet transform (FWT) \cite{beylkin-141} enables fast and
accurate transformation between these two representations. From a
practical perspective, we simply pick the most convenient basis to
compute in, just as one would if using the Fourier basis and the fast
Fourier transform. However, the disjoint support of
multi-wavelets enables local adaptive refinement, which is not
possible in the global Fourier basis. A third equivalent representation is the
value of the function at the Gauss-Legendre quadrature points within
each box of the mesh, which facilitates certain operations (such as
multiplication of functions).

We can use a similar approach to understand how to turn smoothness
into sparsity in applying operators. Let $P^n$ and $Q^n$ be
the projectors onto $V^n$ and $W^n$, respectively. Then, by
construction, $P^{n+1} = P^n + Q^n$. Let $T$ be the kernel of an
operator we wish to apply and let us assume that level $n$ is adequate
to resolve both the input and output functions. The 
representation of the operator at level $n$ is
\begin{subequations}
\begin{align}
  T^n & = P^n T P^n \\
      & = \left(P^{n-1} + Q^{n-1} \right) T \left(P^{n-1} + Q^{n-1} \right) \nonumber \\
      & = T^{n-1} + Q^{n-1} T Q^{n-1} + Q^{n-1} T P^{n-1} + P^{n-1} T Q^{n-1} \nonumber \\
      & = T^{0} + \sum_{m=0}^{n-1} \left( A^m + B^m + C^m \right),
\end{align}
\end{subequations}
where we define $A^m = Q^m T Q^m$, $B^m = Q^m T P^m$, and $C^m = P^m T
Q^m$. In this so-called non-standard form \cite{beylkin-354}, $T^0$
acts only upon scaling functions at the coarsest level whereas the operators 
applied at finer levels all involve wavelets.

For a demonstrative example of the computational significance of this
approach, let us examine convolution with the Poisson kernel \cite{alpert-149,harrison-103,harrison-11587}, $T = | \vec{x} - \vec{y} |^{-1}$. Consider
the operator $A^m$, which connects wavelets at level $m$ to other
wavelets at level $m$. By construction,
the first $k$ moments of the wavelets vanish; thus, the multipole
expansion of $T$ tells us that the matrix elements of $A^m$ decay as
$r^{-2k-1}$ (\textit{e.g.},\ if $k=8$, which is routine, they decay as
$r^{-17}$). In other words, even though $T$ is dense, $A^m$ is very
sparse to any finite precision. A similar approach shows that $B^m$
and $C^m$ decay as $r^{-k-1}$. Translationally invariant operators
yield Toeplitz representations, which are very important for practical
computation due to the greatly reduced burden of computing and storing the
operator.

\subsection{Separated representations}
\label{sec:numerics:sepreps}

The approach described above is not efficient beyond one dimension
because the cost of na\"ively applying an operator grows as $O\left(
(2k)^{2D} \log(\epsilon^{-1})^D \right)$ with similarly excessive
memory requirements, where $D$ is the dimensionality and $\epsilon$ is
the desired precision.  However, for many physically important
operators we can construct \cite{beylkin-235,harrison-103,harrison-11587} either global or local
(via numerically generated low-rank factorizations of the full
operator) separated representations; that is, we can approximate, with
controlled precision, the full operator as a practically short sum of
products of operators that separately apply in each dimension. This
reduces the cost of application to $O \left( M
(2k)^{D+1} \log(\epsilon^{-1})^D \right)$ and the memory footprint to
$O \left( M D (2k)^2 \log(\epsilon^{-1}) \right)$ where $M$ is the
number of terms in the sum. Powerful techniques for generating these separated
forms include quadrature of integral identities \cite{harrison-103} and
numerical techniques developed by Beylkin and Monz\'on \cite{beylkin-131}. For example, numerical quadrature to evaluate the
integral
\begin{align}
  r^{-1} & = \frac{2}{\sqrt{\pi}}\int_{-\infty}^{\infty} \mathrm{d}s e^{-r^{2}e^{2s}+s} \nonumber \\
\intertext{yields compact representations
of $r^{-1}$ as a sum over Gaussians,}
         & = \sum_{\mu=1}^M { c_{\mu} e^{-t_{ \mu } r^2}} + O\left(\epsilon r^{-1} \right);
\end{align}
in which the number of terms ($M$) scales only as $O\left(\left( \log R \right)  \left( \log \epsilon^{-1} \right) \right)$, where $[1,R]$ is the range of validity 
and $\epsilon$ is the relative precision.

\section{Applications: Programming with MADNESS}
\label{sec:examples}
Having discussed the mathematical underpinnings of MADNESS, we briefly
examine some samples of numerical programming with MADNESS.
Many additional examples, including quantum chemistry and other
applications, are provided in the MADNESS code repository (see the
directories \texttt{src/apps/} and \texttt{src/examples/} in the code
\cite{madness}).

\subsection{Example: Getting Started with MADNESS}
Our first example aims to evaluate the integral (trace), 2-norm, and electrostatic self-energy of a Gaussian function in three dimensions.
\begin{subequations}
\begin{align}
g(\vec{x}) & = e^{-|\vec{x}|^2} \\
\int \mathrm{d}^3 \vec{x} \; g(\vec{x}) & = \pi^{3/2} \doteq 5.5683279 \\
\left( \int \mathrm{d}^3 \vec{x} \; g(\vec{x})^2 \right)^{1/2} & = (\pi/2)^{3/4} \doteq 1.403104 \\
\int \mathrm{d}^3 \vec{x} \; g(\vec{x}) \int \mathrm{d}^3 \vec{y} \; |\vec{x}-\vec{y}|^{-1} g(\vec{y}) & = \pi^{5/2} 2^{1/2} \doteq 24.739429 \label{eq:gaussian-interaction}
\end{align}
\label{eq:gaussian}
\end{subequations}
Listing \ref{code:gaussian} shows the entire MADNESS source code for this task, which we now walk through.

\lstset{caption={Code for evaluating the integrals in Eq.\ \eqref{eq:gaussian}.} This code can be found in \texttt{src/examples/gaussian.cc.}, label=code:gaussian}
\begin{lstlisting}
#include <madness/mra/mra.h>
#include <madness/mra/operator.h>
using namespace madness;

double gaussian(const coord_3d& r) {
    double x=r[0], y=r[1], z=r[2];
    return exp(-(x*x + y*y + z*z));
}

int main(int argc, char** argv) {
    initialize(argc, argv);
    World world(SafeMPI::COMM_WORLD);
    startup(world,argc,argv);

    FunctionDefaults<3>::set_cubic_cell(-6.0,6.0);
    real_convolution_3d op = CoulombOperator(world, 1e-4, 1e-6);
    real_function_3d g = real_factory_3d(world).f(gaussian);
    print(g.trace(), g.norm2(), inner(g,op(g)));

    finalize();
    return 0;
}
\end{lstlisting}
Initially ignoring the boiler-plate code, there are only four lines of
significance.
\begin{itemize}
\item Line 15 defines the computational volume ($[-6,6]^3$).
\item Line 16 constructs the Coulomb operator; in actuality, a
      separated Gaussian representation of the Green's function for
      Poisson's equation with free-space boundary conditions, as
      described in \S\ref{sec:numerics:sepreps}. The two numerical
      values in the constructor indicate the finest length scale to
      resolve and the desired accuracy of the operator, respectively.
\item Line 17 projects the analytic function (lines 5-8) into
the adaptive numerical basis using the default precision ($\epsilon=10^{-4}$)
and wavelet order ($k=6$).
\item Line 18 prints the results. 
\end{itemize}
The integral in \eqref{eq:gaussian-interaction} (the last printed value) is
computed as the inner product
(\texttt{inner}) of the Gaussian function (\texttt{g}) and its
convolution with the Green's function (\texttt{op(g)}). Printed
results agree with the exact values to six significant figures, which
is more than might be expected from the default precision of
$10^{-4}$. This is a common observation and is likely due to the
projection into the numerical basis oversampling to guarantee
the requested precision.

The main program commences (line 11) by initializing the MADNESS
parallel computing environment (discussed in \S\ref{sec:computation}),
which includes initializing MPI (if it has not yet been initialized).
Next, line 12 creates a \texttt{World} object that encapsulates an MPI
intra-communicator; this plays the same role as an MPI communicator,
but provides the much more powerful capabilities of the MADNESS
runtime.  Subsequently, line 13 initializes the MADNESS numerical
environment. The program concludes by terminating the parallel
computing environment (line 20), which also finalizes MPI if the call
to \texttt{initialize()} initialized MPI.

Finally, the MADNESS framework provides support for common
mathematical and computational processes. For example, algorithms are
implemented for (i) matrix-free iterative solvers (such as
GMRES \cite{saad-856} and BiCGStab \cite{van-der-vorst-631}), (ii) converting
between the multi-wavelet basis and a Fourier basis
\cite{jia-5870}, (iii) solving non-linear systems of equations \cite{harrison-328}, and
(iv) outputting MADNESS functions for visualization with various
tools, including VisIt \cite{visit}, OpenDX \cite{opendx}, and ParaView \cite{henderson}.

\subsection{Molecular electronic structure}
One of the initial targets when developing MADNESS was quantum
chemistry, where electronic structure problems for atoms and molecules are
solved using the density functional theory (DFT)
or the Hartree-Fock methods \cite{harrison-103, harrison-11587, yanai-6680}.
Either of these approaches requires solving a set of nonlinear partial
differential equations, which, for time-independent systems, are
typically cast into an eigenvalue problem,
\begin{equation}
\label{eq:ks}
\left(-\frac{1}{2}\nabla^2 + V(\vec{x})\right) \psi(\vec{x}) = E \psi(\vec{x}).
\end{equation}
$V(\vec{x})$ is the potential energy for an electron at position
$\vec{x}$ and is itself dependent on $\psi(\vec{x})$ (thus the
nonlinearity). The eigenvalues are interpreted as energies of
molecular states and the eigenfunctions as molecular orbitals.

The standard approach for solving Eq.\ \eqref{eq:ks} expands
$\psi(\vec{x})$ in a predefined, fixed set of basis functions, thereby converting
Eq.\ \eqref{eq:ks} into a generalized matrix equation. Our MADNESS
implementation follows a different approach \cite{kalos-1791} to
utilize the computational strengths of MADNESS and to 
avoid the numerical problems associated with derivative operators
on deeply refined meshes.
We recast Eq.\ \eqref{eq:ks} as an
integral equation,
\begin{equation}
\label{eq:ls-integral}
\psi(\vec{x}) = -2 \int \mathrm{d}\vec{x} (-\nabla^2 - 2 E)^{-1} V(\vec{x}) \psi(\vec{x}),
\end{equation}
which can be solved as a fixed point iteration to the lowest
eigenvalue.  
The integral kernel in
Eq.\ \eqref{eq:ls-integral} (which assumes free-space boundary conditions),
\begin{equation}
\label{eq:bshkernel}
G(\vec{x},\vec{x}') = \frac{1}{4 \pi} \frac{e^{-\sqrt{-2E}}}{|\vec{x}-\vec{x}'|}
\end{equation}
not only incorporates the correct long-range asymptotic form for
bound-state (\textit{i.e.},\ $E<0$) molecular orbitals, but it can be
efficiently applied using the separated representations discussed
in \S\ref{sec:numerics:sepreps}.
Capabilities of the molecular electronic structure code include energies for a
wide range of functionals (including hybrids) and for many-body methods (second-order M{\o}ller-Plesset in six dimensions \cite{bischoff-104103, bischoff-114106} and configuration interaction \cite{kottmann-asap}), localized orbitals for reduced
scaling, forces \cite{yanai-2866}, solvation \cite{fosso-tande-179}, and
linear-response for excited states and dynamic polarizabilities \cite{sekino-034111,yanai-413}.

\subsection{Nuclear physics}

Nuclear physics is another successful application for MADNESS \cite{fann-012080,pei-012035,pei-024317,pei-064306}.
Nuclear density functional theory (NDFT) extends the
molecular DFT above to describe the complex superfluid many-fermion and boson
system. A critical difference of
NDFT is the use of two-component and four-component complex orbitals with
particle spins and the appearance of resonance and continuum states. For
example, the two-component complex wave-functions, $u$ and $v$, and an
associated pairing potential, $\Delta$, extend the one-component orbital
($\psi$) in the molecular DFT to NDFT. These modifications
ultimately result in the Hartree-Fock-Bogoliubov (HFB) equations,
\begin{equation}
\label{eq:hfb}
\begin{bmatrix}
h_{\uparrow} -\lambda_{\uparrow} & \Delta \\
\Delta^{*} & h_{\downarrow} -\lambda_{\downarrow}
\end{bmatrix}
\begin{bmatrix}
u_i \\
v_i
\end{bmatrix}
= E_i
\begin{bmatrix}
u_i \\
v_i
\end{bmatrix},
\end{equation}
where $h_{\uparrow\downarrow}$ is a single particle Hamiltonian for
the given component and $\lambda_{\uparrow\downarrow}$ is the chemical
potential for that component.  For
the ultracold fermionic gas in the BEC-BCS (Bose-Einstein condensate - Bardeen-Cooper-Schriffer) crossover simulation, the oscillatory
Fulde-Ferrell-Larkin-Ovchinnikov phase was found to have
oscillating pairing gaps such that
\begin{equation}
\label{eq:fflo}
h_{\uparrow\downarrow} = -\frac{1}{2} \nabla \cdot \left[\nabla \alpha_{\uparrow\downarrow}(r)\right] + V_{\uparrow\downarrow}(r).
\end{equation}
$\alpha_{\uparrow\downarrow}$ is the kinetic energy density and $V_{\uparrow\downarrow}$ is the trapping potential. The HFB equations are solved with a complex version of the scheme used by the molecular density functional application with additional geometric constraints; Eq.\ \eqref{eq:hfb} is transformed into a coupled set of integral equations, to which Eq.\ \eqref{eq:bshkernel} is applied.

\subsection{Atomic and molecular physics}

The above applications are all time-independent (or introduce
time-dependence only through response theory). In \cite{vence-033403}
MADNESS was used by physicists interested in the effects of
non-perturbative, few-cycle laser pulses on atoms and molecules.
This research prompted us to develop numerically robust and accurate
techniques for evolving quantum wavepackets (and more general systems
of partial differential and integral equations) forward in time. The
familiar challenges of time evolution are exacerbated by adaptive
refinement and truncation of the multi-wavelet basis. Basis truncation
introduces high-frequency noise, as do the inevitable discontinuities
between adjacent sub-domains. This noise must be removed from the
simulation for accuracy and efficiency. Unfortunately, we cannot
casually accept that frequencies beyond some band limit (cut-off) will
be propagated inaccurately; for example, the wave function of the
time-dependent Schr\"odinger equation (TDSE) depends upon accurate
phase information. Our solution was to choose a band limit, propagate
exactly below the band limit, and filter frequencies above it ---
essentially as if we were computing on a globally very fine, uniform
mesh but still retaining the advantages of adaptive refinement.

The TDSE was solved \cite{vence-033403}
for hydrogenic atoms and other systems
subjected to a strong, attosecond laser field, and the
photoionization cross section and other observables were computed. The electron
behavior was modeled by a three-dimensional wave function, $\Psi(\vec{x},t)$,
which, within the dipole approximation and in atomic units, evolves according
to the TDSE,
\begin{equation}
  i \frac{d\Psi(\vec{x},t)}{dt} = \left( -\frac{1}{2} \nabla^2 + V(\vec{x}) + \vec{E}(t) \cdot{} \vec{x} \right) \Psi(\vec{x},t),
\end{equation}
where $\vec{E}$ is the electric field and $V$ the atomic potential. The initial
state, $\Psi(\vec{x},0)$, is the ground state; that is, the solution of Eq.\
\eqref{eq:ks} with the smallest $E$.

The most efficient evolution was performed with a fourth-order accurate, gradient-corrected exponential propagator \cite{chin-7338}
that is only slightly more expensive than the second-order accurate Trotter approximation. Crucial for efficient
and accurate application was the realization that, while the kernel of the
free-space propagator, $$(2 \pi i t)^{-1/2} \exp(-|\vec{x}|^2/2it),$$ 
is infinite in extent and highly oscillatory, its combination with a projector
onto a finite bandlimit is compact and also bandlimited.

\section{Computational Framework: The MADNESS Parallel Runtime}
\label{sec:computation}
\begin{figure}
\resizebox{4in}{!}{\includegraphics{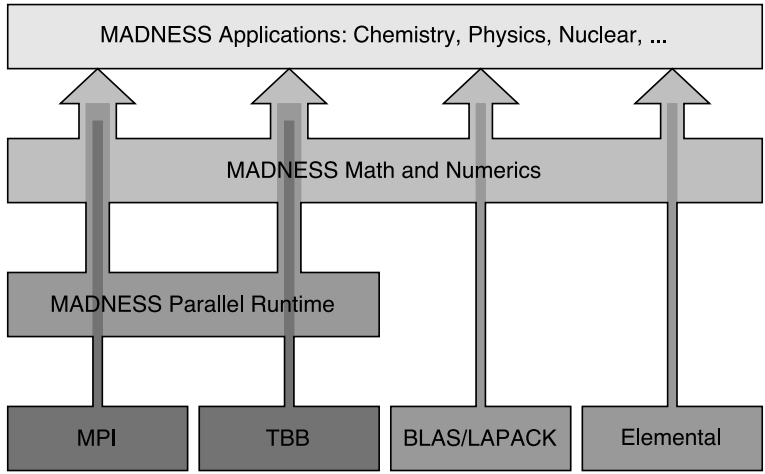}}
\caption{\label{fig:arch} The library structure of MADNESS and its dependencies. Each of these sub components has its own public application programing interface (API) that may be used both for development inside MADNESS as well as in external software packages.}
\end{figure}

The MADNESS parallel runtime, like the numerical libraries of MADNESS, is an intuitive interface that allows the user to compose algorithms as a dynamically scheduled set of tasks that operate on objects in global namespaces.
This high-level approach to parallel programming offers greater composability than that of the traditional
Message Passing Interface (MPI) and explicit thread programming (POSIX or C++ threads). 
The key runtime features include the use of
\begin{remunerate}
\item {\em global namespaces} for building applications that are composed of (global) objects interacting via remote methods;
\item {\em tasks} as first-class entities for fine-grained work decomposition;
\item {\em futures} \cite{baker-55} for expressing dependencies between the tasks.
\end{remunerate}
These features permit the programmer to deal less with the explicit low-level details of computation and data partitioning (\textit{e.g.}, message passing between processes with disjoint data or threads operating on shared data) and focus more on high-level concepts central to the scientific domain of interest. An algorithm properly composed within the MADNESS runtime will (partially) hide costs of data communication and be tolerant to load imbalance due to dynamic, data-driven work scheduling.

Many of the individual concepts appear in other packages, for example, in Cilk \cite{blumofe-207, blumofe-55}, Charm++ \cite{kale-overview, kale-91, kale-tech-report}, Intel Threading Building Blocks (TBB) \cite{tbb}, and other projects
including ACE \cite{ace-url, schmidt} and PGAS languages \cite{yelick-24}. Some of these features are starting to make their way into mainstream programming languages. C++, for example, has included task-based concurrency with asynchronous function calls and futures since 2011 (note that
MADNESS futures, unlike C++ futures, can refer not only to results of local tasks, but also to those of
{\em remote} tasks).

For the sake of portability the MADNESS parallel runtime is implemented in C++ (2011 standard), MPI-1 and POSIX threads; all non-portable software prerequisites, such as Intel TBB, are optional. In the most common scenario, each node executes a single MPI process
composed of the main application thread. The MADNESS runtime spawns
a remote method invocation (RMI) thread and a pool of computational threads
managed by a task queue; these threads are not directly accessible to the user.
The main application thread runs sequentially through the main program and submits tasks 
(that can themselves generate more tasks) to
the local and/or remote task queue.\footnote{The main thread can also act as a computational thread by executing tasks in the task queue while waiting for an event (\textit{e.g.}, while waiting inside a barrier).
} The output of each task is encapsulated in a future that can serve as input to other tasks.
Once its input futures are available (assigned), the task queue schedules the tasks for execution. Because MADNESS futures live in a global namespace, assigning a future can involve
data movement to a remote process; this process is facilitated by the RMI thread in the receiving process.
When available, MADNESS can use the task queue provided by the Intel TBB library; MADNESS defaults to its own task queue implemented in terms of POSIX threads.

By supporting the straightforward composition of a distributed task-based application,
the MADNESS runtime allows programmers to focus on exposing concurrency by
decomposing data and computation and/or by optimizing load balance. The low-level
details of thread scheduling and synchronization, data movement, and
communication hiding are automated. However, the MADNESS runtime provides access 
to enough low-level details of the architecture (\textit{e.g.}, process rank)
that the placement of data and computation can be directly controlled by the
user. This allows tight orchestration of algorithms with complex data flows,
such as in the systolic loops used for computing localized electronic states
in molecular electronic structure. These abilities, as provided by the MADNESS
runtime, are also used by \texttt{TiledArray} \cite{CalvinValeevSC15Paper} (a
framework for block-sparse tensor computations) to hide communication costs and
withstand load imbalances in handling block-sparse data.

In addition to the above core capabilities, the MADNESS parallel runtime also
provides higher-level constructs built on these features. One such feature is a
distributed-memory, associative container used to store the functions and
operators in the MADNESS numerical library. This container automates the
placement and computation of distributed data. Other features include parallel
algorithms (\textit{e.g.}, parallel for-each) and task-based collective
operations. Many of these components interoperate via futures and, therefore, fit
naturally within the task-based framework. For example, the task-based all-reduce
algorithm, which is analogous to the \texttt{MPI\_Allreduce} function, uses
futures as both the input and output of the function, allowing a seamless
connection between local computation tasks, collective communication, and
remote-computation tasks.

\section{Summary}
\label{sec:summary}
Since the introduction and initial successful science application of
MADNESS, it received an R\&D 100 Award in 2011 \cite{madness-r&d100},
and there have been at least three independent implementations of the
associated numerical methods in Japan \cite{sekino-355},
Norway \cite{frediani-1143}, and the United States \cite{valeevprivate}.
MADNESS itself is now thriving as an open-source, community project
with financial support from multiple sources.  Central to people's
interest is the emphasis on a high-level of composition while
maintaining guarantees of accuracy and speed. It is not that MADNESS
is necessarily the fastest code for any particular problem, but rather
that a user working on tractable problems will get the right
answer with modest effort and without having to place unnecessary emphasis
on computational details.  However, the caveat of ``tractability'' is
non-trivial --- user attention is still needed to regularize some
singularities, and the fast algorithms within MADNESS (notably the
integral convolution kernels) do not yet include scattering operators,
which is a subject of current research.  Similarly, efficient
computation is presently limited to simple domains and, as shown
in \cite{reuter-1}, cumbersome techniques are presently necessary to
accommodate even simple interior boundary conditions.  Nevertheless,
the numerical techniques have demonstrated their worth in a broad range of
physically interesting applications.  Similarly, the MADNESS parallel
runtime is being successfuly used for petascale computations
independent of the numerical layer \cite{CalvinValeevSC15Paper},
illustrating the power and utility of the massively threaded,
task-based approach to computation.

\bibliographystyle{siam}
\bibliography{refs}
\end{document}